\documentclass[sigconf, review = false, nonacm = true]{acmart} 

\usepackage{booktabs}
\usepackage{multirow}
\usepackage{verbatim}
\usepackage{graphicx}
\usepackage{amsmath}
\usepackage{algorithm} 
\usepackage{bm}

\newcommand{\specialcell}[2][l]{%
  \begin{tabular}[#1]{@{}l@{}}#2\end{tabular}}
 \newcommand\modelAbbrv{M2H\xspace}
\newcommand\model{M2H model\xspace}
\newcommand\modelCapitalized{M2H Model\xspace}

\newcommand\modelNoWhitespace{M2H model\xspace}
\newcommand\ie{i.\,e.\xspace}
\newcommand\eg{e.\,g.\xspace}

\usepackage{enumitem}
\usepackage{xspace}
\usepackage{xparse}
\usepackage{algpseudocode}

\usepackage{amsmath}

\DeclareMathOperator*{\argmin}{arg\,min}

\newcommand{\norm}[1]{\left\lVert#1\right\rVert}

\usepackage{etoolbox}
\robustify\bfseries

\AtBeginDocument{%
  \providecommand\BibTeX{{%
    \normalfont B\kern-0.5em{\scshape i\kern-0.25em b}\kern-0.8em\TeX}}}

\begin{document}

\title{Predicting COVID-19 Spread from Large-Scale Mobility Data}

\author{Amray Schwabe}
\email{schwabea@ethz.ch}
\orcid{0000-0001-7890-6687}
\affiliation{%
  \institution{ETH Zürich}
  \city{Zürich}
  \country{Switzerland}
  \postcode{8092}
}

\author{Joel Persson}
\email{jpersson@ethz.ch}
\orcid{0000-0002-2831-1535}
\affiliation{%
  \institution{ETH Zürich}
  \city{Zürich}
  \country{Switzerland}
  \postcode{8092}
}

\author{Stefan Feuerriegel}
\email{sfeuerriegel@ethz.ch}
\orcid{0000-0001-7856-8729}
\affiliation{%
  \institution{ETH Zürich}
  \city{Zürich}
  \country{Switzerland}
  \postcode{8092}
}

\begin{abstract}
To manage the COVID-19 epidemic effectively, decision-makers in public health need accurate forecasts of case numbers. A potential near real-time predictor of future case numbers is human mobility; however, research on the predictive power of mobility is lacking. To fill this gap, we introduce a novel model for epidemic forecasting based on mobility data, called mobility marked Hawkes model. The proposed model consists of three components: (1)~A Hawkes process captures the transmission dynamics of infectious diseases. (2)~A mark modulates the rate of infections, thus accounting for how the reproduction number $R$ varies across space and time. The mark is modeled using a regularized Poisson regression based on mobility covariates. (3)~A correction procedure incorporates new cases seeded by people traveling between regions. Our model was evaluated on the COVID-19 epidemic in Switzerland. Specifically, we used mobility data from February through April 2020, amounting to approximately 1.5 billion trips. Trip counts were derived from large-scale telecommunication data, \ie, cell phone pings from the \emph{Swisscom} network, the largest telecommunication provider in Switzerland. We compared our model against various state-of-the-art baselines in terms of out-of-sample root mean squared error. We found that our model outperformed the baselines by 15.52\,\%. The improvement was consistently achieved across different forecast horizons between 5 and 21 days. In addition, we assessed the predictive power of conventional point of interest data, confirming that telecommunication data is superior. To the best of our knowledge, our work is the first to predict the spread of COVID-19 from telecommunication data. Altogether, our work contributes to previous research by developing a scalable early warning system for decision-makers in public health tasked with controlling the spread of infectious diseases.
\end{abstract}



\maketitle

\section{Introduction}

COVID-19 has evolved into a global pandemic, which, as of January 30, 2021, has been responsible for more than 100 million reported cases \cite{who_coronavirus_2021}. The spread of COVID-19 continues to pose a serious threat not only to public health but also to healthcare institutions and economies as a whole. To control the spread of COVID-19, decision-makers in public health need accurate predictions of the future case numbers. This allows for early interventions and, on top of that, is important for near real-time resource planning, so that hospital capacities are not exceeded \cite{ferguson_report_2020}. 

To make epidemic forecasts in near real-time, leading indicators are needed \cite{grantz_use_2020}. Such leading indicators should predict the spread of infectious diseases and, to be of practical value, should be scalable. In epidemiology, spatio-temporal data on mobility has been suggested as a leading indicator that could enable epidemic forecasting as infectious disease are spread through human-human transmissions \cite{desai_stopping_2020}. In previous work, epidemic forecasting using mobility has been evaluated for dengue fever \cite{kim_modeling_2019}, Ebola \cite{kelly_real-time_2019}, and cholera \cite{bengtsson_using_2015}. Yet, in the case of COVID-19, the out-of-sample predictive power of mobility data for epidemic forecasting is subject to research.

Human mobility can be monitored using a variety of methods. One approach is to quantify mobility using point of interest (POI) data obtained from check-ins \cite{chang_mobility_2021, benzell_rationing_2020, dave_when_2021} or smartphone location logs\cite{chinazzi_effect_2020,li_substantial_2020, bonaccorsi_economic_2020,chang_mobility_2021}. An alternative is to extract mobility patterns from telecommunication data \cite{persson_monitoring_2021}. Telecommunication data is regarded to provide a more comprehensive coverage of a country's population \cite{oliver_mobile_2020, grantz_use_2020}. In particular, telecommunication data have, unlike POI data, the advantage of capturing routine activities that might not otherwise be recorded. Telecommunication data originates from cell phone pings and thus track \emph{all} SIM card movements, regardless of which hard- or software is used by the phone. Motivated by this, several research commentaries have discussed the potential benefits of telecommunication data for monitoring the spread of COVID-19 \cite{oliver_mobile_2020, grantz_use_2020}. However, research actually evaluating the predictive power of telecommunication data for forecasting the spread of COVID-19 is still lacking. This represents the objective and novelty of the present work.

\textbf{Proposed model (\modelAbbrv)\footnote{Both code and data will be made available via \url{https://github.com/amrayschwabe/PredictingCovidSpreadMobility}.}:} 
We propose a novel mobility marked Hawkes model (\modelAbbrv) for epidemic forecasting based on mobility data. The \model consists of three components: (1)~A Hawkes process captures the transmission dynamics of infectious diseases. The benefit of a Hawkes process is that each event (infection) may trigger further offspring events, thus reflecting the transmission dynamics of infectious diseases. In our model, each region is represented by a different instantiation of the Hawkes process. (2)~A mark modulates the rate of infections, thus accounting for spatio-temporal variation in the reproduction number $R$. The mark is modeled using a regularized Poisson regression based on mobility covariates extracted from telecommunication data. (3)~A correction procedure incorporates new cases seeded by people traveling between regions. The model is thus semi-mechanistic in the sense that its components are based on transmission dynamics and human behavior. 

\textbf{Findings:} Our proposed \model was evaluated with large-scale mobility data from Switzerland. We derived mobility based on telecommunication metadata from \emph{Swisscom}, the largest telecommunication provider in Switzerland. Mobility data from February through April 2020 was used, consisting of approximately 1.5 billion trips. The \model was evaluated over forecast horizons ranging from 5 to 21 days. In particular, we compared the root mean squared error (RMSE) against state-of-the-art baselines \cite{chiang_hawkes_2020, wieczorek_neural_2020, persson_monitoring_2021}. Overall, we found that the RMSE of our model outperformed the baselines by 15.52\,\%. In addition, we re-estimated our model with POI data (\ie, Google Community Mobility Reports \cite{google_llc_google_2020}) instead of telecommunication data. This decreased the prediction performance by 10.57\,\%, thus confirming that the use of telecommunication data as done in our study is superior.

\textbf{Contributions:}
Our contributions are as follows:
\begin{enumerate}
    \item We present a novel mobility marked Hawkes model for epidemic forecasting. Compared to previous research, our model captures the effects of between-region transmissions and incorporates the effects of mobility based on telecommunication data. 
    \item We outperform state-of-the-art prediction models when forecasting COVID-19 spread by 15.52\,\%. The superior performance is consistently demonstrated for forecast horizons between 5 and 21 days.
    \item We establish the predictive power of large-scale telecommunication data for forecasting COVID-19 spread. We further show that mobility flows from telecommunication data provide higher predictive power than POI data (i.e., Google Community Mobility Reports).
\end{enumerate}
As implications for practice, we help decision-makers in public health by providing a tool for near real-time epidemic forecasting and thus enable better management of the COVID-19 epidemic.

\section{Background}

\subsection{Modeling the Spread of COVID-19}

Predicting COVID-19 spread is a challenging task. Previous research has used a range of explanatory models, which describe the spreading dynamics retrospectively (e.\,g., \cite{persson_monitoring_2021, banholzer_impact_2020}). In contrast, few models have been developed (and evaluated) for forecasting. Epidemic forecasting is, however, crucial in practice as such forecasts inform decision-makers in public health (e.g., when timing interventions or opening strategies, and when managing the utilization of healthcare resources \cite{ferguson_report_2020}). In the following, we provide a brief review of predictive models for epidemic forecasting.

Generally, models for epidemic forecasting can be distinguished into stochastic and mechanistic models. Stochastic models mainly rely on the predictive power of data, while mechanistic models explicitly utilize epidemiological theory and empirical evidence. An example of a stochastic model often used in the early stages of disease spread is an exponential model \cite{bertozzi_challenges_2020}. Another example of a stochastic model is a Gaussian model, which is used to predict the peak of COVID-19 cases \cite{schlickeiser_gaussian_2020}. Both models predict future cases solely from past cases. Recent efforts have incorporated additional predictors. In the case of COVID-19, predictors have been derived from behavioral data such as visitor counts for POIs or call data, and used been used in regression models  \cite{rostami-tabar_forecasting_2021} and machine learning models \cite{wieczorek_neural_2020, kapoor_examining_2020}. However, machine learning models are of limited use in emerging epidemics due to their need for large historical training data. Yet, in epidemic forecasting, such large historical training data are typically not available. 

One type of mechanistic model is the compartmental model. In a compartmental model, each individual belongs to some compartment (e.g., susceptible, infected, recovered) and has a certain probability of transitioning to another compartment. The probability of a transition can either be derived from case data or modeled as being dependent on additional predictors. Both approaches have been used in previous research on COVID-19 \cite{pei_differential_2020, chang_mobility_2021, bertozzi_challenges_2020, chinazzi_effect_2020, miller_mobility_2020, zhao_prediction_2020,qian_when_2020}. However, a downside of compartmental models is, that they cannot fully account for uncertainty. For instance, a fixed value has to be used for the incubation time.

In between stochastic and mechanistic models are point processes, such as the Hawkes process \cite{hawkes_cluster_1974}. The self-exciting structure of point processes enables the modeling of how events (i.e., infections) trigger further offspring events. This property reflects human-human transmission in which infected individuals transmit disease to other people \cite{bertozzi_challenges_2020}. Under some conditions, Hawkes processes have been shown to provide equivalent estimates to compartmental models \cite{rizoiu_sir-hawkes_2018}. Moreover, Hawkes processes can be extended to include a distribution over the incubation time. Due to these benefits, Hawkes processes have previously been used for modeling COVID-19 spread \cite{bertozzi_challenges_2020, chiang_hawkes_2020, mohler_analyzing_2020}. Yet, to the best of our knowledge, we are the first to incorporate mobility flows from telecommunication data.

\subsection{Mobility Data}
\label{sec:mobilityData}

Mobility data has been extensively used for explaining urban phenomena. Previous applications include predicting crime patterns \cite{kadar_leveraging_2020} and detecting the impact of social investments \cite{zhou_cultural_2017}. On an urban scale, previous research has also explored the predictive power of mobility data in forecasting chronic diseases such as cancer or diabetes \cite{wang_predicting_2018}.

In the case of COVID-19, previous work has empirically confirmed the association between mobility and COVID-19 cases based on explanatory analysis \cite{xiong_mobile_2020, persson_monitoring_2021}, while predictive modeling is scarce. To obtain mobility data, one approach is to back-trace human movements from POI data (static data). POI data provide visitor counts to specific locations. Such counts can be collected manually through check-ins or automatically (e.g., Google Community Mobility Reports \cite{google_llc_google_2020}).

Alternatively, mobility data can be derived from movements across space and time (dynamic data) resulting in origin-destination pairs. In contrast to static data, dynamic data have the advantage of tracing the underlying trajectory and, therefore, are likely to provide a more comprehensive description of mobility. Dynamic data might thus be able to predict COVID-19 spread better than static data.

\subsection{Hawkes Processes}

A Hawkes process is a self-exciting point process characterized by an intensity function depending on all previous events \cite{hawkes_cluster_1974}. The intensity function describes processes where one event is likely to trigger a series of offspring events. Previous work models disease spread as a na\"ive Hawkes process \cite{kim_modeling_2019, kelly_real-time_2019, schoenberg_recursive_2019, chiang_hawkes_2020}, yet none of these works leverages telecommunication data to model the spatio-temporal spreading dynamics.

A specific variant of a Hawkes process is the marked Hawkes process. A so-called ``mark'' measures the ''impact`` of an event. A marked Hawkes process has been used in applications such as forecasting earthquakes \cite{ogata_statistical_1988}. In this work, we develop a novel, tailored version of the marked Hawkes process for epidemic forecasting based on mobility data.

\textbf{Research gap:} Mobility has been suggested as a leading indicator of the epidemic, which can be inferred from telecommunication data. However, research assessing the predictive power of telecommunication data is lacking. To close this gap, we introduce a novel mobility marked Hawkes process for epidemic forecasting.

\section{The Proposed Model}

Let $\mathcal I = \{i=1,\ldots,N\}$ be a finite set of mutually exclusive regions (e.g., states). Further, let $t=1,\ldots,T$ denote a fixed number of days. Our goal is to predict the daily count of new cases $y_{it} \in \mathbb N_{\geq 0}$ in each region over a forecast horizon. We will do so by modelling the observed trajectories of reported cases, $\{y_{i1},y_{i2},\ldots,y_{iT}\}^N_{i=1}$, as realizations from underlying discrete-time stochastic processes $\{Y_i(t) \colon t=1,\ldots,T\}^N_{i=1}$.

\subsection{Model Components}

Our \model consists of three components. A discrete-time \textbf{Hawkes process} for the underlying transmission dynamics, a modeled \textbf{mark} for the reproduction rate, and a \textbf{correction for between-region mobility}, incorporating new cases seeded by infected people traveling between regions. Later, the mark accommodates within-region mobility, while the correction is responsible for incorporating between-region mobility. In the following, we describe each component. 

\textbf{(1) Hawkes process:} Here we describe the standard Hawkes process \cite{hawkes_cluster_1974}, which we then adapt to our discrete-time setting. A Hawkes process is a self-exciting point process. It models disease spread as a time series of cases where every case can produce offspring cases. Additionally, so-called immigrant cases are triggered by external sources and contribute to disease spread. A standard Hawkes process is defined by the intensity function
\begin{equation}
    \label{eq:continuous_hawkes}
    \lambda(s) = \lambda_0 + \sum_{s_j < s} \phi(s-s_j).
\end{equation}
The intensity function $\lambda(s)$ represents the expected number of cases at time $s$. It depends on the background rate $\lambda_0$ (rate of immigrant cases) and $\sum_{s_j < s} \phi(s-s_j)$, the rate of offspring cases. This depends on all past cases $s_j$ where $\phi(s-s_j)$ regulates the change that a case at time $s_j$ has on the intensity function at time $s$. In our case, the kernel function $\phi$ models the distribution of incubation times. Note that the Hawkes process is a particular case of a non-homogeneous Poisson process, in which the intensity is stochastic and explicitly depends on previous events through the kernel function.

The standard Hawkes process is not directly applicable to modeling infectious diseases, since case numbers are usually available on a daily basis and, therefore, in discrete time. Hence, instead of a standard Hawkes process, we change the definition to one in discrete time:
\begin{equation}
    \label{eq:discrete_hawkes}
    \lambda(t) = \lambda_0 + \sum_{t_j < t} y_{t_j} \phi(t-t_j),
\end{equation}
where $y_{t_j}$ is the sum of cases reported on day $t_j$. Instead of summing over all past events $s_j < s$ occurring in continuous time, we now sum over all past days $t_j < t$ where, on day $t_j$, $y_{t_j}$ events happened. Note that, even in this formulation, the events themselves need not be in discrete time, but are simply aggregated over discrete time intervals corresponding to days. Accordingly, $\lambda(t)$ now denotes the intensity function on day $t$. 

\textbf{(2) Mark:} We introduce a mark $R$ to account for the spatio-temporal variation in the number of offspring cases per case, commonly called the reproduction rate. The mark is a function of covariates and denotes the ``impact'' of an event. For instance, a mark of two corresponds to two offspring cases being caused by every past infection. The mark is incorporated into the Hawkes process via
\begin{equation}
\label{eq:marked_hawles}
    \lambda(t) = \lambda_0 + \sum_{t_j < t} y_{t_j} R(\bm x_t) \phi(t-t_j),
\end{equation}
where $R(\bm x_t)$ is the realization of the mark given some covariates $\bm x_t$, which we will introduce later. In comparison to Eq.~\ref{eq:discrete_hawkes}, the influence of a case at time $t_j$ on the intensity function $\lambda(t)$ may now not only depend on the time difference $t-t_j$, but also on the day $t_j$ itself. The resulting model is a marked Hawkes process in discrete-time. We model the mark with a regularized Poisson regression, explained in the following.

We estimate the mark in region $i$ on day $t$ with regularized Poisson regression. We chose a Poisson regression for two reasons: (1)~Poisson regression is robust against heteroskedastic error terms, whereas estimation with ordinary least squares is not and would give inconsistent parameter estimates \cite{winkelmann_econometric_2008}. (2)~Empirical research in epidemiology has estimated the reproduction rate as Poisson distributed \cite{liu_reproductive_2020, you_estimation_2020}. 

Two types of covariates are included in the regression: a vector $\bm m_{it} = [m_{it1},\ldots,m_{itz}]$ of mobility covariates that vary by region and day, and $w_{it} \in \mathbb R$ as a summary measure of the weather in region $i$ at day $t$. In our case, the mobility covariates are the count of within-region trips for different purposes (commuting vs. non-commuting) and using different modes of transport (highway, road, and train). Within-region mobility is added in the Poisson regression, while between-region mobility is part of the correction in component~(3). For the weather covariate, we use the maximum temperature per region and day.

Let $\bm x_{it} = [\bm m_{it},w_{it}]$ and $\bm \theta = [\bm \beta, \psi]$ be the covariate and parameter vectors, respectively. The regression model is 
\begin{equation}
    \label{eq:poisson}
    \mathbb E[R(\bm X_{it}; \beta_0, \bm \theta)|\bm X_{it} = \bm x_{it}; \beta_0, \bm \theta]
    = 
    \exp( \beta_0 + \bm \theta^\top \bm x_{it}).
\end{equation}
The parameters are estimated by fitting the regression on the inferred reproduction rate. We use a lasso regularization \cite{tibshirani_regression_1996} to avoid overfitting and facilitate model interpretation by shrinking non-predictive covariates towards zero. Note that the regularization term is not part of the regression model itself but added as a constraint in the estimation. Details are provided in Sec.~\ref{sec:estimation}.

\textbf{(3) Correction for between-region mobility:} We introduce a correction procedure that incorporates new cases seeded by people traveling between regions. The procedure is motivated by the following observation: An offspring case in any region can be caused by one of two things: (1)~a past case in the same region, or (2)~a past case in any other region. The standard Hawkes process only accounts for the effect of~(1) while implicitly attributing the effect of~(2) to the background rate $\lambda_0$. However, $\lambda_0$ is restricted to be constant over time with effects on the intensity function independent of the mark. This is problematic for modeling infectious diseases as the rate of immigrant cases likely depends on (a)~the number of cases in other regions, (b)~the inflow of individuals from other regions, and (c)~the reproduction rate within the same region, all of which may change over time. To capture these dynamic dependencies, we use the following three-step correction procedure:

\begin{enumerate}
    \item Estimate the number of infected individuals $n_i(t)$ traveling into region $i$ on day $t$ from any other region with
    \begin{equation}
        \label{eq:nc}
        \hat n_i(t) \coloneqq \sum_{i' \in \mathcal I \setminus \{i\}} m_t(i',i) \frac{y_{i't}}{q_{i'}},
    \end{equation}
    where $m_t(i',i)$ is the count of trips from region $i'$ to region $i$ on day $t$, $y_{i't}$ is the sum of infected individuals at time $t$ in any region $i'$, and $q_{i'}$ is the population size of region $i'$.
    \item Multiply $n_i(t)$ by a weighting factor $\alpha$ to allow for a differential effect of spreading dynamics between regions compared to within regions. 
    \item Replace the background rate $\lambda_0$ with $\alpha n_i(t)$ and add it to $y_{i't}$ in the marked Hawkes process.
\end{enumerate}
Altogether, the procedure captures the increase in the intensity function due to transmission of the disease caused by between-region trips of infected people. Note that, in our model, there is no need to include the background rate $\lambda_0$ (which is otherwise present in standard Hawkes processes). The reason is that immigrant cases are already captured by the added term $\alpha n_i(t)$.

\subsection{The \texorpdfstring{\modelCapitalized}{LG}}

Combining the above components gives our \model. Specifically, its intensity function for region $i$ and day $t$ is given by
\begin{align}
    \lambda_i(t) 
    &= \sum_{t_j < t} (y_{it_j}+\alpha \hat n_i(t_j)) \cdot R(\bm x_{it}; \beta_0, \bm \theta)\cdot \phi(t-t_j),
\end{align}
where $\hat n_i(t_j)$ and $R(\bm x_{it}; \beta_0, \bm \theta)$ are given by Eq.~\ref{eq:nc} and  Eq.~\ref{eq:poisson}, respectively. The model is a marked Hawkes process, in which for each region, disease transmissions occur due to people within the region and those traveling into the region. The future spread is modulated by the reproduction number (i.e., mark), which itself is modeled as a function of the amount and type of mobility and weather. We model the kernel function $\phi$ as the incubation time distribution by adding the constraint
\begin{equation}
    \sum_{l \in \mathbb N^+} \phi(l) = 1.
\end{equation}
This defines $\phi$ as a distribution function. Now, $\phi$ only determines the timing of offspring cases and thus separably identifies effects of the incubation period to those of the reproduction number. Note that a standard Hawkes process does not model these effects.

\subsection{Model Estimation}
\label{sec:estimation}
We estimate our \model by developing an expectation-maximization (EM) algorithm \cite{dempster_maximum_1977} that is tailored to our task. This approach is computationally efficient and thus enables fast model estimation and prediction. The algorithm is given in Alg.~\ref{alg:em}.

\begin{algorithm}[H]
\small
	\caption{EM algorithm for the \model} 
	\label{alg:em}
	\begin{algorithmic}[1]
	\Procedure{EM}{$\alpha$, $\xi$, $\phi$, $\{y_{it}, \bm x_{it}, \hat n_i(t) \colon \forall i,t\}$}
	\State initialize $R(\bm x_{it};\beta_0, \bm \theta) = 1$ $\forall i,t$
	\While {$\norm{\Delta \bm \theta }_1 >$ \text{tol} }
		 \State \textit{Expectation step:}
			\For {${i=1,2,\ldots,N}$}
			    \For {$\forall$ $t, t' = 1,...,T$ where $t < t'$ }
			   \State $p_i(t,t') = \frac{R(\bm x_{it};\beta_0, \bm \theta) \cdot \phi(t'-t)}{\lambda_i(t')}$
			    \EndFor
			   \For {$t = 1,...,T$}
			   \State $r_{it} = \sum_{t' > t} p_i(t,t') \cdot y_{it'}$
			   \EndFor
			\EndFor
		\State \textit{Maximization step:}
		\State $\hat{\bm \theta}, \hat{\bm \beta}= \argmin_{\beta_0,\bm \theta} - \frac{1}{NT} \log \mathscr L(\beta_0, \bm \theta | \bm X, \bm R) + \xi
		\norm{\bm \theta^\top}_1$
	\EndWhile
	\EndProcedure
	\end{algorithmic}
\end{algorithm}

The EM algorithm proceeds as follows: It takes the weight $\alpha$, the lasso penalty $\xi$, and the kernel function $\phi$ as given and initializes the estimated reproduction rates $R(\bm x_{it})$. In the expectation step, then estimates the probability $p_i(t,t')$ that one case on a day $t'$ is caused by one in day $t$, for all days in the data. Then, the reproduction rate $r_{it}$ is calculated from the obtained probabilities and the number of cases. This procedure is repeated for all regions. 

In the maximization step, the parameters $\bm \theta$ of the Poisson regression are estimated by minimizing the negative log-likelihood function subject to the lasso regularization. The regression model is specified as a generalized linear model with log-link. The log-likelihood function of $\bm \theta$ given the design matrix $\bm X$ (with elements $\bm x_{it}$) and the reproduction rate matrix $\bm R$ (with elements $r_{it}$) is 
\begin{equation*}
    \log \mathscr L(\beta_0, \bm \theta | \bm X, \bm R) 
    = \sum^N_{i=1} \sum^T_{t=1}
    \Big(
    r_{it}(\beta_0 + \bm \theta^\top \bm x_{it}) - \exp(\beta_0 + \bm \theta^\top \bm x_{it})
    \Big).
\end{equation*}
The expectation and maximization steps were repeated until convergence (until $\beta_0$ and $\bm \theta$ vary by less than $10^{-5}$). All covariates in the regression were standardized to aid convergence. 

Informed by research on incubation times \cite{lauer_incubation_2020}, we model $\phi$ as a discretized Gamma$(5.807, 1.055)$ distribution. Specifically, let $F$ be the cumulative distribution function (CDF) of the Gamma distribution. Then $\phi(l) = F(l+0.5) - F(l-0.5)$ for $l \in \mathbb N^+$, with $\phi(1) \coloneqq F(1.5)$ to cover the probability mass on the interval $[0,0.5]$. 

To estimate the weight $\alpha$ and lasso penalty $\xi$, we conduct a grid search to minimize the prediction error of the reported number of cases in a leave-one-out (here, region) cross-validation setting. The intervals for $\alpha$ and $\xi$ were $[0,3]$ and $[0,20]$, respectively.

\subsection{Forecasting}

To predict the number of future COVID-19 cases in region $i$ at $\delta$ days ahead, we proceed as follows: 
\begin{enumerate}
    \item Starting at day $t$, estimate the reproduction rate so far using the values of the covariates until day $t$ and the estimated parameters of the regularized Poisson regression.
    \item An intensity function estimate $\hat \lambda_i(t)$ is then obtained by evaluating the model at the parameter estimates $(\hat \beta_0, \hat{\bm \theta})$.
    \item Draw $Y_{it} \sim\text{Poisson}(\hat \lambda_i(t))$ as a prediction for the number of cases in region $i$ at day $t$.
    \item Repeat steps (1)--(3) for day $t+d$ with $d=1,\ldots,\delta$.
\end{enumerate}
Steps (1)--(4) are repeated ten times for each region $i=1,\ldots,N$, resulting in ten predictions of future disease spread for all regions. The ten different predictions reflect the uncertainty in the future number of cases in a given region. 

\subsection{Model Variants}
\label{sec:model_variants}

We tested several variants of our model as part of an ablation study (see Sec.~\ref{sec:ablationStudy}). This allowed us to assess which parts of the model improve prediction performance.
\begin{enumerate}
    \item \textbf{\model without regularization:} We evaluated a variant of our \model without regularization to test whether the penalty is necessary.
    \item \textbf{\model without weather covariate:} Weather could be an important determinant of COVID-19 spread, and, to test this, we evaluated our model against one without the weather covariate.
    \item \textbf{\model without correction term:} To evaluate if our proposed correction procedure enhances prediction performance, we also considered a model in which it is replaced with the background rate. 
    \item \textbf{\model with between-region mobility covariates in the mark:} To evaluate if our correction procedure sufficiently accounts for between-region mobility, we added the corresponding covariates to the regression model for $R$.
    \item \textbf{\model with demographic covariates:} Following prior literature \cite{chiang_hawkes_2020}, we tested if prediction performance is increased by adding demographic covariates (here: population density, population size, and percentage of city population).
    \item \textbf{\model with POI data:} To evaluate if telecommunication data offers higher predictive power, we re-estimated our model with POI data derived from the Google Community Reports \cite{google_llc_google_2020}. We did not apply our correction for between-region travel, as POI data does not provide any information about between-region mobility.
\end{enumerate}

\section{Setting}

\subsection{Data}
\label{sec:data}

The study period was Feb 24, 2020--April 26, 2020. This period covers the first wave of COVID-19 in Switzerland, starting at the date of the first case and ending when the number of new cases again approached zero.

\textbf{Case data:} In Switzerland, COVID-19 cases are reported at the state level (called ``cantons''). We model each of the 26 cantons as a separate region. In our results, we refer to regions as ``cantons''. The daily number of COVID-19 cases per canton was obtained from the Federal Office of Public Health of the Swiss Confederation (BAG) \cite{federal_office_of_public_health_coronavirus_2021}. No additional preprocessing is applied to the case data.

\textbf{Mobility data:} Telecommunication data were used to infer mobility of SIM-based devices. In our work, telecommunication data was obtained from a collaboration with \emph{Swisscom}, the largest telecommunication provider in Switzerland. The data collection and processing procedures are compliant with data privacy laws in Switzerland. Ethics approval from ETH Zurich was obtained (2020-N-41). 

Data collection was conducted in collaboration with \emph{Swisscom} as follows. First, routine signal exchanges (``pings'') between mobile devices and antennas were recorded. The data comprises \underline{all} pings, regardless of the mobile service provider. Second, a triangulation algorithm was used to approximate the location (longitude, latitude) of each SIM card \cite{kafsi_quantifying_2019}. The accuracy of the triangulation has been tested and amounts to a median error of 132 meters \cite{kafsi_quantifying_2019}. Location data was recorded at a resolution of approx. 5 minutes. This ensures that our data capture micro-level movements of individuals of high granularity. Third, the movements were used to compute trips. A trip is registered when a SIM card moves between two different postcode areas and the destination location has been static for at least 20 minutes. Fourth, trips were classified by their mode of transport (``train'', ``road'', and ``highway'') and their purpose (``commuter'' vs. ``non-commuter''). The mode of trips is determined via the positions of antennas along train rails, highways, and roads. If a trip uses multiple modes of transport, it is classified as the mode that covered the longest distance. ``Commuter'' and ``non-commuter'' trips are differentiated based on extracted home and work locations of individuals. Details are reported in the Swisscom Mobility Insights \cite{swisscom_swisscom_2021}. Fifth, trips were aggregated for each day and region. This results in an origin-destination matrix of mobility flows. We then derive within- and between-region mobility covariates. The within-region mobility covariate is calculated as the reduction in within-region trips at a given day relative to a reference period (here: the average daily trip count during the two weeks before the first Swiss COVID-19 case). The between-region covariates are for each region calculated as the total number of trips into the region per every other region.

\textbf{Additional predictors:} Weather data were retrieved per day and canton from the COVID-19 open data repository\footnote{https://github.com/googleCloudPlatform/covid-19-open-data, retrieved: January 10, 2020}. The daily maximum temperature was used as a predictor. For one model variant, we included covariates for the sum of incoming trips per canton. These were obtained for each canton by calculating the change in the sum of incoming trips relative to the reference period. For another model variant,  predictors included socio-demographic statistics (canton population, population density, percentage of city population, age group distribution). These were obtained from the Swiss Federal Statistical Office \cite{office_population_2021}. 

\subsection{Summary Statistics}

\begin{figure}
  \centering
  \begin{tabular}{@{}c@{}}
  \label{before}
    \includegraphics[width=.90\linewidth]{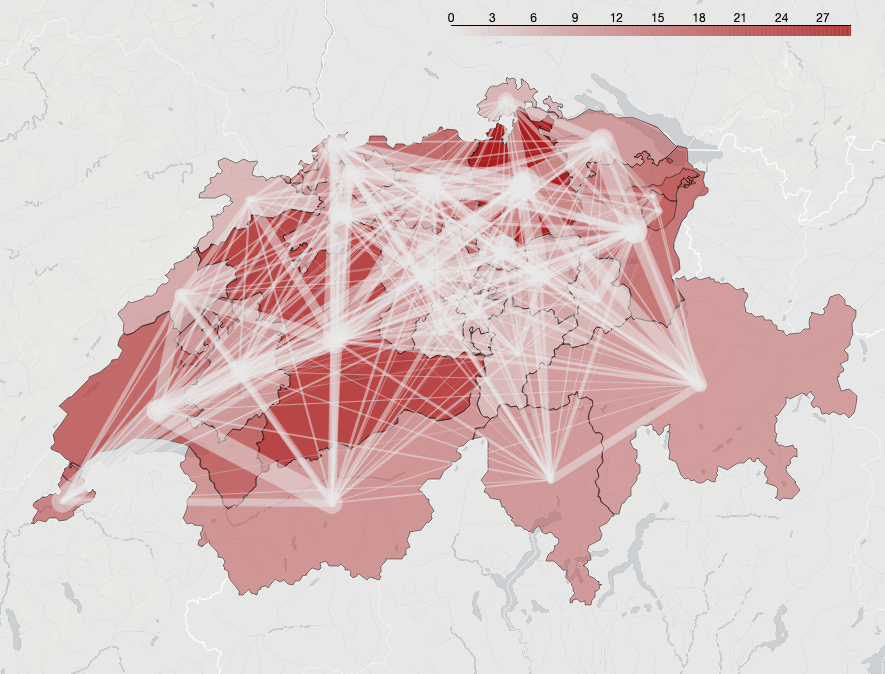} \\[\abovecaptionskip]
    \small (a) Mobility flows between cantons on February 21, 2020.
  \end{tabular}

  \vspace{\floatsep}

  \begin{tabular}{@{}c@{}}
  \label{after}
    \includegraphics[width=.90\linewidth]{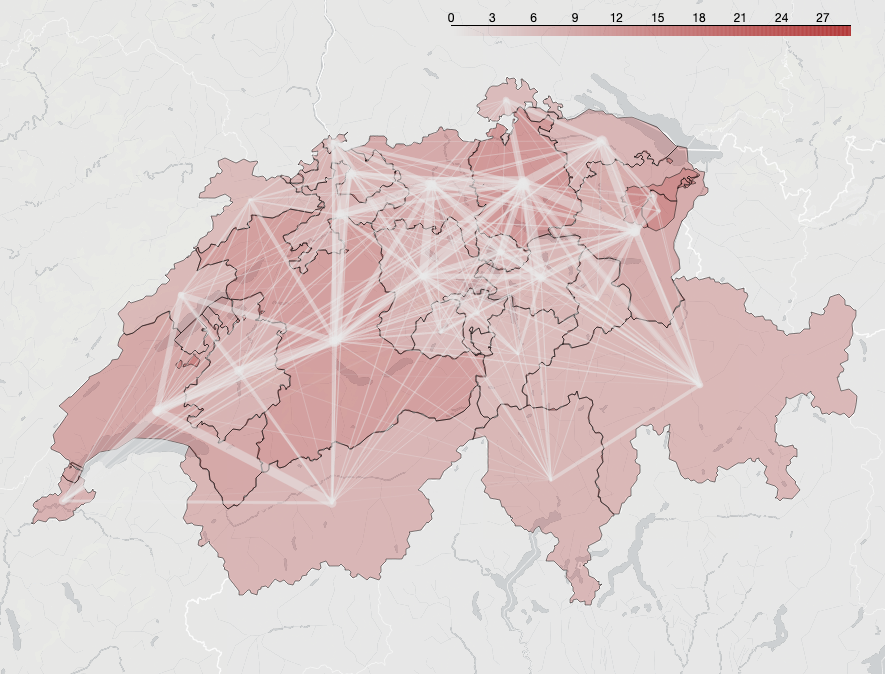} \\[\abovecaptionskip]
    \small (b) Mobility flows between cantons on March 29, 2020.
  \end{tabular}

  \caption[Caption for LOF]{Comparison of mobility flows (i.e., origin-destination matrix) on two example dates before COVID-19 (a) and after (b) restrictions have been imposed. Line width denotes volume of between-region trips; color denotes volume of within-region trips (in millions).}
\label{fig:mobilityFlows}
\end{figure}

\textbf{Case data:} During our study period, 29,148 COVID-19 cases were registered in Switzerland. COVID-19 spread started on February 24, 2020, with a maximum of 1455 people tested positive on March 23, 2020. COVID-19 spread has varied considerably between regions (i.e., cantons), with only 20 cases registered for the canton of Appenzell Innerrhoden (AI) until the end of our study period (April 26, 2020), but 5308 cases for the canton of Vaud (VD).

\textbf{Mobility data:} Our study period covers approximately 1.5 billion trips. Following the first lockdown in Spring 2020, overall mobility in Switzerland dropped by approximately 50\,\%. The mobility flows before and after the lockdown are shown in Fig.~\ref{fig:mobilityFlows}.

\subsection{Evaluation Framework}

Predictions were evaluated across different forecast horizons (5--21 days ahead). A leave-one-region-out validation was used to test the out-of-sample predictive performance of our model. This ensures that the model does not overfit a specific region. Two performance metrics are reported: the root mean squared error (RMSE) and the mean absolute error (MAE). The former is a statistical measure of model fit whereas the latter quantifies the deviation in the number of cases and is thus of relevance for decision-making in public health. 

\subsection{Baselines}

Our model is compared against the following baselines:
\begin{enumerate}
  \item \textbf{Negative binomial regression} \cite{persson_monitoring_2021}:
    The negative binomial regression model predicts future case numbers by using lagged mobility indices, time trends, and region-specific factors, and accounts for overdispersion in case numbers.
    \item \textbf{ARMAX model:} The ARMAX($p,q$) model is a common baseline for time-series forecasting. It models the daily case count as being dependent on that of the previous $p$ days, the previous $q$ days' estimation errors, and additional covariates. This model should thereby capture time dependence in the cases and prediction errors. The model orders $p = 6, q = 1$ were estimated via a grid search ($p, q \in 1,2..,10$).
    \item \textbf{Random forest:} A random forest is a non-parametric machine learning model that forecasts case counts from covariates using an ensemble of decision trees. This baseline is chosen due to its flexibility in data-scarce settings and handling non-linearity.
    \item \textbf{Neural network} \cite{wieczorek_neural_2020}: An artificial neural network is used to predict case counts based on past cases and additional covariates. The weather covariate did not increase performance and was thus discarded. The architecture and hyperparameter tuning (via grid search) was done analogous to \cite{wieczorek_neural_2020}.
    \item \textbf{SEIR mobility model \cite{kermack_contribution_1927}}: A SEIR model predicts future case numbers from past infections in a deterministic manner. Here, we allow the reproduction rate to be modulated by mobility by multiplying the weighted change in mobility with the initial reproduction rate. 
    \item \textbf{Na\"ive marked Hawkes process} \cite{chiang_hawkes_2020}: In the na\"ive marked Hawkes process, the mark is determined from demographic covariates and within-region mobility. Different from our \modelNoWhitespace, regularization for $R$ is not included and the static background rate is used instead of our correction procedure. However, to capture between-region heterogeneity, socio-demographic covariates are additionally included.
\end{enumerate}
We also experimented with a \textbf{graph neural network} from \cite{kapoor_examining_2020}, yet, for our problem setup, found it inapplicable. Hence, this model was discarded from our comparison. Unless otherwise stated. all models were fed with the same data as our proposed \modelNoWhitespace.

\section{Results}

\subsection{Overall Prediction Performance}

Table \ref{tab:results} reports the macro-averaged prediction performance of the models across the different regions and forecast horizons. Our proposed \model achieves an RMSE of 13.20 and an MAE of 11.07. For comparison, the best baseline (the neural network from \cite{wieczorek_neural_2020}) has an RMSE of 15.41 and an MAE of 12.87. Hence, our model improves both RMSE and MAE by 15.52\,\%, and 16.26\,\%, respectively. Note that all modes are evaluated with the same data, and hence, the performance improvement is due to the model itself. 

\begin{table}[h!]
\small
  \caption{Comparison of overall prediction performance (i.e., macro-averaged across regions and forecast horizons).}
  \label{tab:results}
  \begin{tabular}{lll}
    \toprule
   Model & RMSE  & MAE\\
    \midrule
      \specialcell{Negative binomial regression \cite{persson_monitoring_2021}} &  19.39 & 16.77\\
     \addlinespace[3pt]
      \specialcell{ARMAX model} & 28.84 & 25.14\\
     \addlinespace[3pt]
      \specialcell{Random forest} &32.78 & 28.33\\
    \addlinespace[3pt]
     \specialcell{Neural network \cite{wieczorek_neural_2020}} &15.41 & 12.87\\
     \addlinespace[3pt]
     \specialcell{SEIR mobility \cite{kermack_contribution_1927}} & 36.48 & 31.37\\
     \addlinespace[3pt]
     \specialcell{Na\"ive marked Hawkes process \cite{chiang_hawkes_2020}}&22.21 & 18.64\\
     \addlinespace[3pt]
     \specialcell{\textbf{\model} (proposed)}  &\textbf{13.34} & \textbf{11.07}\\
    \bottomrule
    Lower values = better (best values in bold) \\
\end{tabular}
\end{table}

\subsection{Prediction Performance across Forecast Horizons}

\begin{figure}[b]
  \centering
  \includegraphics[width=0.9\linewidth]{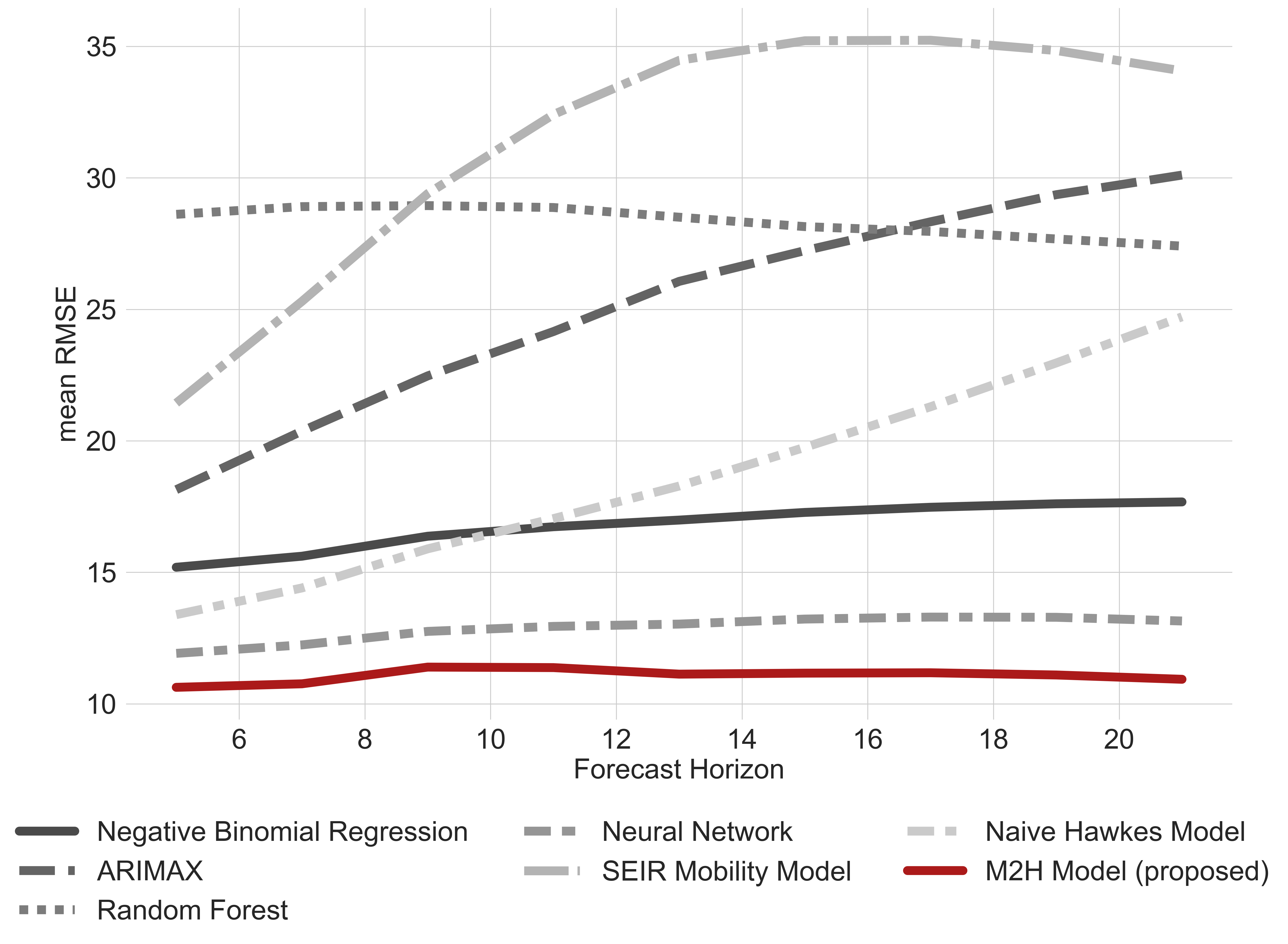}
  \caption{Comparison of prediction error across different forecast horizons macro-averaged by regions.}
  \Description{The prediction error over different forecast horizons of our model versus baseline models.}
  \label{fig:differentHorizons}
\end{figure}

Fig.~\ref{fig:differentHorizons} compares the prediction performance across different forecast horizons. The proposed \model is superior across all baselines and forecast horizons. Moreover, the prediction performance of our model is largely invariant to the forecast horizon. Interestingly, the improvement over the na{\"i}ve Hawkes model increases with longer forecast horizons. The results for the MAE are analogous but omitted due to space constraints. 

\subsection{Actual vs. Predicted Case Counts by Region}

\begin{figure}[H]
  \centering
  \includegraphics[width=0.9\linewidth]{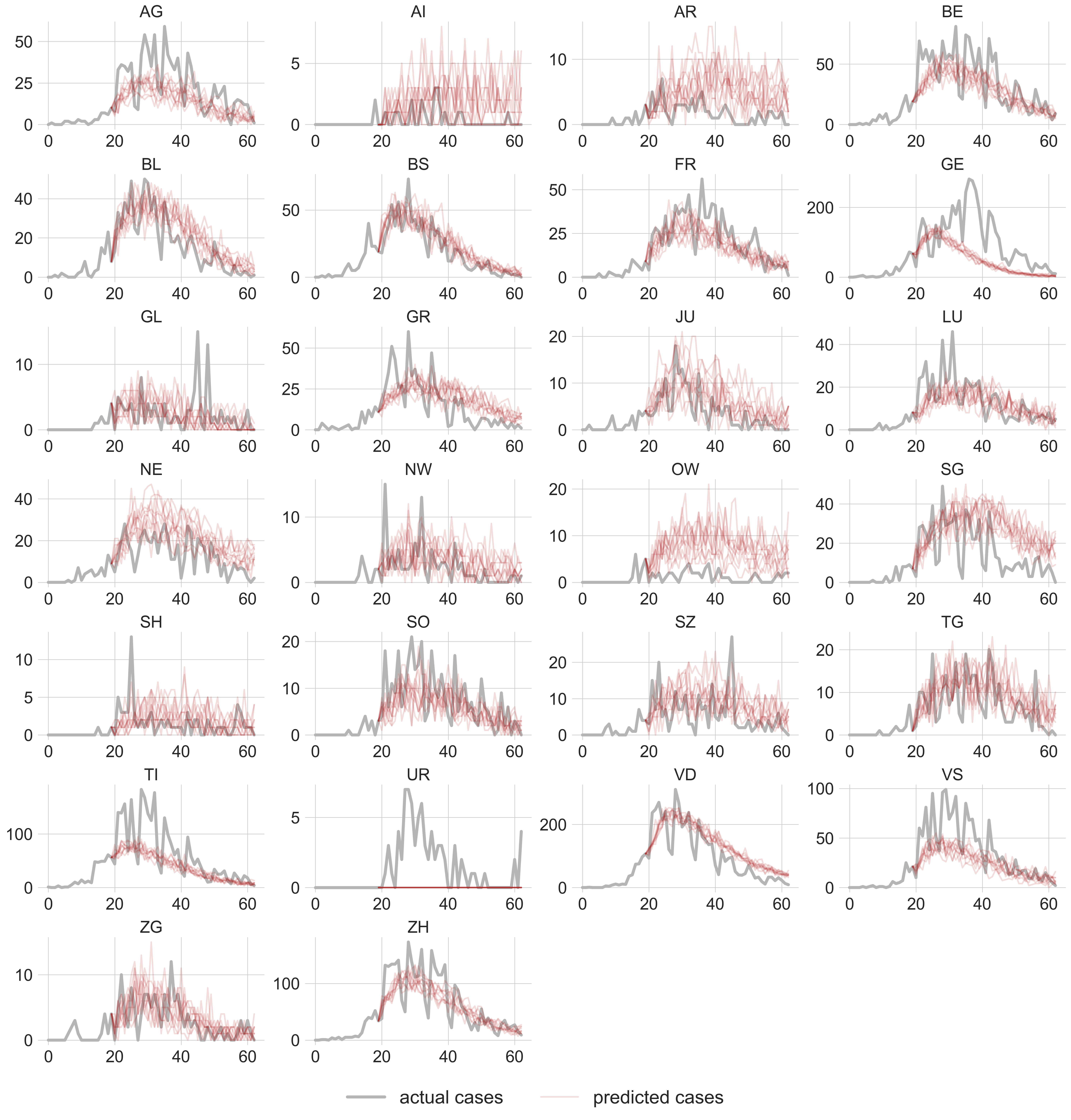}
  \caption{Predictions (gray) vs. actual case counts (red) per region (\ie, per canton). Results are shown for 21-day ahead forecasts and using 10 predictions to indicate uncertainty. The $x$-axis denotes the number of days since the first case in Switzerland, and the $x$-axis denotes the case counts. Names of regions (cantons) are stated using the official two-letter abbreviations. }
  \label{fig:LOCOallCantons20days}
  \Description{Prediction performance over different time horizons}
\end{figure}

We plot predictions of case counts from our model against the actual counts (Fig.~\ref{fig:LOCOallCantons20days}). Here, the variation in the predictions reflects the uncertainty in future cases. Some cantons experienced only a low number of daily cases, oftentimes below 10 (e.g., Appenzell Innerrhoden (AI), Uri (UR), and Obwalden (OW)). The figure shows that our model yields predictions of similar magnitude for these cantons. Other cantons had over 200 daily cases (e.g., Zurich (ZH)). Again, we find that the predictions fit the actual case numbers well.

For the canton of Geneva (GE), predictions appear too conservative. A closer inspection revealed that case counts in Geneva peaked considerably later than in other cantons, and also much later compared to the reduction in mobility. Nonetheless, the predictions of the best baseline (i.e., the neural network from \cite{wieczorek_neural_2020}) were also too conservative. This suggests that the relationship between mobility and cases were different for Geneva as compared to other cantons.

\subsection{Ablation Study}
\label{sec:ablationStudy}

An ablation study presented in Table \ref{tab:results_variations} compares the prediction performance of different variants of our  \model (see Sec.~\ref{sec:model_variants}). All model variants produce predictions inferior to those of the full model, confirming that all parts of the model contribute to its prediction performance. For instance, removing the Lasso regularization reduces the RMSE by 2.62\,\%; removing the correction for between-region mobility reduces the RMSE by 2.62\,\%; and removing the weather covariate reduces the RMSE by 3.15\,\%. Furthermore, adding either demographic information or the incoming trips as covariates did not improve prediction performance. The results are qualitatively the same when examining the change in MAE.  

We further compare the predictive power of static (POI data) vs. dynamic data (mobility flows from telecommunication data). For this purpose, we re-estimate our model with POI based mobility data (here: Google Community Mobility Reports \cite{google_llc_google_2020}). Using POI data increases the RMSE from 13.34 to 14.75, implying a performance drop of 10.57\,\%. A Wilcoxon signed-rank test (cf. \cite{flores_utilization_1989} for its use in forecasting) rejects the null hypothesis that the two models' prediction errors are from the same distribution ($p$-value $< 0.01$). Therefore, the comparison shows that telecommunication data is of higher predictive power.

\begin{table}
\small
  \caption{Ablation study.}
  \label{tab:results_variations}
  \begin{tabular}{lll}
    \toprule
   Model variant & RMSE & MAE \\
    \midrule
     \specialcell{\model w/o regularization}  &13.69 & 11.28\\
     \addlinespace[3pt]
    \specialcell{\model w/o correction for travel} &13.69 & 11.38\\
     \addlinespace[3pt]
     \specialcell{\model w/o weather}  &13.76 & 11.47\\
     \addlinespace[3pt]
     \specialcell{\model with between-region covariates}
     &13.38 & 11.11 \\
     \addlinespace[3pt]
    \specialcell{\model with demographic covariates} &13.56 &11.23\\
     \addlinespace[3pt]
       \specialcell{\model based on POI data} &14.75& 12.32 \\
     \addlinespace[3pt]
     \specialcell{\textbf{\model}}  &\textbf{13.34} &\textbf{11.07}\\
    \bottomrule
    Lower values = better (best values in bold) \\
\end{tabular}
\end{table}

\section{Discussion}

\textbf{Findings:} Our model enables accurate epidemic forecasting. We demonstrate its performance over forecast horizons ranging from 5 to 21 days. Using the same mobility data, the \model outperforms standard baseline models by 15.52\,\%. 

\textbf{Why is the \model superior?} The improved predictive performance of our model can be attributed to its ability to capture infectious disease properties. (1)~In infectious diseases, each case triggers new (offspring) cases. These transmission dynamics are captured by using a Hawkes process. (2)~The rate of infections (i.e., reproduction rate) varies across space and time and is associated with population mobility. The mark models the reproduction rate using spatio-temporal covariates representing within-region mobility and weather. (3)~Between-region mobility leads to spatial spread by seeding new transmission chains in distant areas. This is captured by our correction term. 

\textbf{Model strengths:} Our \model has several benefits for practical applications: First, the model has a parsimonious structure, simplifying the interpretation and avoiding overfitting. Because of this, our model performs well in data-scarce contexts such as epidemic forecasting. Second, our model captures the prediction uncertainty by probabilistic forecasting. Third, it requires minimal hyperparameter tuning. The model is estimated via leave-one-out cross-validation. Here, the weight $\alpha$ and the lasso penalty $\xi$ were obtained via grid search, and the other parameters are estimated with a tailored EM algorithm.

\textbf{Static vs. dynamic data:} Our study demonstrates the effectiveness of mobility flows derived from telecommunication data for epidemic forecasting. The data in our study were obtained from \emph{Swisscom}, the largest telecommunication provider in Switzerland. During the COVID-19 epidemic, the use of telecommunication data for epidemic forecasting has been considered by governments in Switzerland and around the world \cite{busvine_european_2020}. However, empirical evidence concerning its effectiveness for forecasting is lacking. We find that using POI data results in a 10.57\,\% higher RMSE, thus reflecting a substantially inferior predictive power. A reason for this is that telecommunication data have larger coverage; it includes every mobile device independent of the operating system and does not require manual opt-in \cite{buckee_aggregated_2020}. Another advantage of telecommunication data is, they they are measured with high frequency (e. g., daily), thereby enabling regularly updated forecasts as needed by decision-makers. Future research and practice is therefore encouraged to consider the use of telecommunication data for quantifying mobility.

\textbf{Limitations and future work:} Our work is subject to limitations that should be addressed in future research. First, our model was validated on data from the first wave of COVID-19 in Switzerland. Future work could deploy and validate our model in other settings. Note that our model should be applicable in a straightforward manner given the general model structure. Second, telecommunication data cannot be assumed to provide a perfect measure of mobility, as it only captures the movements of individuals with SIM cards. Nevertheless, telecommunication data is argued to be more comprehensive than other sources \cite{buckee_aggregated_2020, grantz_use_2020}. Third, we use various predictors informed by prior literature. The predictive power of other variables could be evaluated in future research.

\textbf{Implications:} Decision-makers in public health need tools for forecasting the spread of epidemics. This helps them in managing interventions (e.g., stay-home orders) and planning scarce healthcare resources (e.g., to avoid over-utilization of ventilators). A challenge is that data on cases are delayed due to incubation times and reporting delays. Hence, scalable tools for near real-time forecasting are needed. For this purpose, research commentaries \cite{bertozzi_challenges_2020,grantz_use_2020, oliver_mobile_2020} have called for research using novel stochastic process models and data on mobility flows. Our model should further be of value in managing other diseases (\eg swine flu, ebola, etc.) and also emerging infectious diseases.

\section{Conclusion}

This work fills a known research gap \cite{grantz_use_2020, oliver_mobile_2020}: leveraging telecommunication data for epidemic forecasting. To achieve this, we propose a novel mobility marked Hawkes process for predicting COVID-19 cases from large-scale telecommunication data. We show that our proposed model outperforms the state-of-the-art in terms of RMSE by 15.52\,\%. The superior performance is achieved consistently across different forecast horizons. Therefore, our prediction model provides a valuable tool for near real-time monitoring of the COVID-19 spread and thus aids decision-makers in public health.

\begin{acks}
We thank Swisscom for the collaboration and the unique data access. One author (Stefan Feuerriegel) is a member of a COVID-19 Working Group of the World Heath Organisation (WHO) and thanks the members for intensive discussions.
\end{acks}

\bibliographystyle{ACM-Reference-Format}
\bibliography{references.bib}


\begin{thebibliography}{50}


\ifx \showCODEN    \undefined \def \showCODEN     #1{\unskip}     \fi
\ifx \showDOI      \undefined \def \showDOI       #1{#1}\fi
\ifx \showISBNx    \undefined \def \showISBNx     #1{\unskip}     \fi
\ifx \showISBNxiii \undefined \def \showISBNxiii  #1{\unskip}     \fi
\ifx \showISSN     \undefined \def \showISSN      #1{\unskip}     \fi
\ifx \showLCCN     \undefined \def \showLCCN      #1{\unskip}     \fi
\ifx \shownote     \undefined \def \shownote      #1{#1}          \fi
\ifx \showarticletitle \undefined \def \showarticletitle #1{#1}   \fi
\ifx \showURL      \undefined \def \showURL       {\relax}        \fi
\providecommand\bibfield[2]{#2}
\providecommand\bibinfo[2]{#2}
\providecommand\natexlab[1]{#1}
\providecommand\showeprint[2][]{arXiv:#2}

\bibitem[\protect\citeauthoryear{Banholzer, van Weenen, Kratzwald, Seeliger,
  Tschernutter, et~al\mbox{.}}{Banholzer et~al\mbox{.}}{2020}]%
        {banholzer_impact_2020}
\bibfield{author}{\bibinfo{person}{Nicolas Banholzer}, \bibinfo{person}{Eva van
  Weenen}, \bibinfo{person}{Bernhard Kratzwald}, \bibinfo{person}{Arne
  Seeliger}, \bibinfo{person}{Daniel Tschernutter}, {et~al\mbox{.}}}
  \bibinfo{year}{2020}\natexlab{}.
\newblock \showarticletitle{Impact of non-pharmaceutical interventions on
  documented cases of {COVID}-19}.
\newblock \bibinfo{journal}{\emph{MedRxiv}} (\bibinfo{year}{2020}).
\newblock


\bibitem[\protect\citeauthoryear{Bengtsson, Gaudart, Lu, Moore, Wetter,
  et~al\mbox{.}}{Bengtsson et~al\mbox{.}}{2015}]%
        {bengtsson_using_2015}
\bibfield{author}{\bibinfo{person}{Linus Bengtsson}, \bibinfo{person}{Jean
  Gaudart}, \bibinfo{person}{Xin Lu}, \bibinfo{person}{Sandra Moore},
  \bibinfo{person}{Erik Wetter}, {et~al\mbox{.}}}
  \bibinfo{year}{2015}\natexlab{}.
\newblock \showarticletitle{Using {Mobile} {Phone} {Data} to {Predict} the
  {Spatial} {Spread} of {Cholera}}.
\newblock \bibinfo{journal}{\emph{Scientific Reports}} \bibinfo{volume}{5},
  \bibinfo{number}{1} (\bibinfo{year}{2015}), \bibinfo{pages}{1--5}.
\newblock


\bibitem[\protect\citeauthoryear{Benzell, Collis, and Nicolaides}{Benzell
  et~al\mbox{.}}{2020}]%
        {benzell_rationing_2020}
\bibfield{author}{\bibinfo{person}{Seth~G. Benzell}, \bibinfo{person}{Avinash
  Collis}, {and} \bibinfo{person}{Christos Nicolaides}.}
  \bibinfo{year}{2020}\natexlab{}.
\newblock \showarticletitle{Rationing social contact during the {COVID}-19
  pandemic: {Transmission} risk and social benefits of {US} locations}.
\newblock \bibinfo{journal}{\emph{PNAS}} \bibinfo{volume}{117},
  \bibinfo{number}{26} (\bibinfo{year}{2020}), \bibinfo{pages}{14642--14644}.
\newblock


\bibitem[\protect\citeauthoryear{Bertozzi, Franco, Mohler, Short, and
  Sledge}{Bertozzi et~al\mbox{.}}{2020}]%
        {bertozzi_challenges_2020}
\bibfield{author}{\bibinfo{person}{Andrea~L. Bertozzi}, \bibinfo{person}{Elisa
  Franco}, \bibinfo{person}{George Mohler}, \bibinfo{person}{Martin~B. Short},
  {and} \bibinfo{person}{Daniel Sledge}.} \bibinfo{year}{2020}\natexlab{}.
\newblock \showarticletitle{The challenges of modeling and forecasting the
  spread of {COVID}-19}.
\newblock \bibinfo{journal}{\emph{PNAS}} \bibinfo{volume}{117},
  \bibinfo{number}{29} (\bibinfo{year}{2020}), \bibinfo{pages}{16732--16738}.
\newblock


\bibitem[\protect\citeauthoryear{Bonaccorsi, Pierri, Cinelli, Flori, Galeazzi,
  et~al\mbox{.}}{Bonaccorsi et~al\mbox{.}}{2020}]%
        {bonaccorsi_economic_2020}
\bibfield{author}{\bibinfo{person}{Giovanni Bonaccorsi},
  \bibinfo{person}{Francesco Pierri}, \bibinfo{person}{Matteo Cinelli},
  \bibinfo{person}{Andrea Flori}, \bibinfo{person}{Alessandro Galeazzi},
  {et~al\mbox{.}}} \bibinfo{year}{2020}\natexlab{}.
\newblock \showarticletitle{Economic and social consequences of human mobility
  restrictions under {COVID}-19}.
\newblock \bibinfo{journal}{\emph{PNAS}} \bibinfo{volume}{117},
  \bibinfo{number}{27} (\bibinfo{year}{2020}), \bibinfo{pages}{15530--15535}.
\newblock


\bibitem[\protect\citeauthoryear{Buckee, Balsari, Chan, Crosas, Dominici,
  et~al\mbox{.}}{Buckee et~al\mbox{.}}{2020}]%
        {buckee_aggregated_2020}
\bibfield{author}{\bibinfo{person}{Caroline~O. Buckee},
  \bibinfo{person}{Satchit Balsari}, \bibinfo{person}{Jennifer Chan},
  \bibinfo{person}{Mercè Crosas}, \bibinfo{person}{Francesca Dominici},
  {et~al\mbox{.}}} \bibinfo{year}{2020}\natexlab{}.
\newblock \showarticletitle{Aggregated mobility data could help fight
  {COVID}-19}.
\newblock \bibinfo{journal}{\emph{Science}} \bibinfo{volume}{368},
  \bibinfo{number}{6487} (\bibinfo{date}{April} \bibinfo{year}{2020}),
  \bibinfo{pages}{145--146}.
\newblock


\bibitem[\protect\citeauthoryear{Busvine}{Busvine}{2020}]%
        {busvine_european_2020}
\bibfield{author}{\bibinfo{person}{Elvira~Pollina Busvine, Douglas}.}
  \bibinfo{year}{2020}\natexlab{}.
\newblock \showarticletitle{European mobile operators share data for
  coronavirus fight}.
\newblock \bibinfo{journal}{\emph{Reuters}} (\bibinfo{date}{March}
  \bibinfo{year}{2020}).
\newblock
\urldef\tempurl%
\url{https://www.reuters.com/article/us-health-coronavirus-europe-telecoms-idUSKBN2152C2}
\showURL{%
\tempurl}


\bibitem[\protect\citeauthoryear{Chang, Pierson, Koh, Gerardin, Redbird,
  et~al\mbox{.}}{Chang et~al\mbox{.}}{2021}]%
        {chang_mobility_2021}
\bibfield{author}{\bibinfo{person}{Serina Chang}, \bibinfo{person}{Emma
  Pierson}, \bibinfo{person}{Pang~Wei Koh}, \bibinfo{person}{Jaline Gerardin},
  \bibinfo{person}{Beth Redbird}, {et~al\mbox{.}}}
  \bibinfo{year}{2021}\natexlab{}.
\newblock \showarticletitle{Mobility network models of {COVID}-19 explain
  inequities and inform reopening}.
\newblock \bibinfo{journal}{\emph{Nature}} \bibinfo{number}{589}
  (\bibinfo{year}{2021}), \bibinfo{pages}{82--87}.
\newblock


\bibitem[\protect\citeauthoryear{Chiang, Liu, and Mohler}{Chiang
  et~al\mbox{.}}{2020}]%
        {chiang_hawkes_2020}
\bibfield{author}{\bibinfo{person}{Wen-Hao Chiang}, \bibinfo{person}{Xueying
  Liu}, {and} \bibinfo{person}{George Mohler}.}
  \bibinfo{year}{2020}\natexlab{}.
\newblock \showarticletitle{Hawkes process modeling of {COVID}-19 with mobility
  leading indicators and spatial covariates}.
\newblock \bibinfo{journal}{\emph{medRxiv}} (\bibinfo{year}{2020}).
\newblock


\bibitem[\protect\citeauthoryear{Chinazzi, Davis, Ajelli, Gioannini, Litvinova,
  Merler, et~al\mbox{.}}{Chinazzi et~al\mbox{.}}{2020}]%
        {chinazzi_effect_2020}
\bibfield{author}{\bibinfo{person}{Matteo Chinazzi},
  \bibinfo{person}{Jessica~T. Davis}, \bibinfo{person}{Marco Ajelli},
  \bibinfo{person}{Corrado Gioannini}, \bibinfo{person}{Maria Litvinova},
  \bibinfo{person}{Stefano Merler}, {et~al\mbox{.}}}
  \bibinfo{year}{2020}\natexlab{}.
\newblock \showarticletitle{The effect of travel restrictions on the spread of
  the 2019 novel coronavirus ({COVID}-19) outbreak}.
\newblock \bibinfo{journal}{\emph{Science}} \bibinfo{volume}{368},
  \bibinfo{number}{6489} (\bibinfo{year}{2020}), \bibinfo{pages}{395--400}.
\newblock


\bibitem[\protect\citeauthoryear{Dave, Friedson, Matsuzawa, and Sabia}{Dave
  et~al\mbox{.}}{2021}]%
        {dave_when_2021}
\bibfield{author}{\bibinfo{person}{Dhaval Dave}, \bibinfo{person}{Andrew~I.
  Friedson}, \bibinfo{person}{Kyutaro Matsuzawa}, {and}
  \bibinfo{person}{Joseph~J. Sabia}.} \bibinfo{year}{2021}\natexlab{}.
\newblock \showarticletitle{When {Do} {Shelter}‐{In}‐{Place} {Orders}
  {Fight} {COVID}‐19 {Best}? {Policy} {Heterogenity} {Across} {States} and
  {Adoption} {Time}}.
\newblock \bibinfo{journal}{\emph{Economic Inquiry}} \bibinfo{volume}{59},
  \bibinfo{number}{1} (\bibinfo{date}{Jan.} \bibinfo{year}{2021}),
  \bibinfo{pages}{29--52}.
\newblock


\bibitem[\protect\citeauthoryear{Dempster, Laird, and Rubin}{Dempster
  et~al\mbox{.}}{1977}]%
        {dempster_maximum_1977}
\bibfield{author}{\bibinfo{person}{Arthur~P. Dempster}, \bibinfo{person}{Nan~M.
  Laird}, {and} \bibinfo{person}{Donald~B. Rubin}.}
  \bibinfo{year}{1977}\natexlab{}.
\newblock \showarticletitle{Maximum {Likelihood} from {Incomplete} {Data} {Via}
  the {EM} {Algorithm}}.
\newblock \bibinfo{journal}{\emph{J R Stat Soc Series B Stat Methodol}}
  \bibinfo{volume}{39}, \bibinfo{number}{1} (\bibinfo{year}{1977}),
  \bibinfo{pages}{1--22}.
\newblock


\bibitem[\protect\citeauthoryear{Desai and Patel}{Desai and Patel}{2020}]%
        {desai_stopping_2020}
\bibfield{author}{\bibinfo{person}{Angel~N. Desai} {and} \bibinfo{person}{Payal
  Patel}.} \bibinfo{year}{2020}\natexlab{}.
\newblock \showarticletitle{Stopping the {Spread} of {COVID}-19}.
\newblock \bibinfo{journal}{\emph{JAMA}} \bibinfo{volume}{323},
  \bibinfo{number}{15} (\bibinfo{year}{2020}), \bibinfo{pages}{1516--1516}.
\newblock


\bibitem[\protect\citeauthoryear{Ferguson, Laydon, Nedjati~Gilani, Imai,
  Ainslie, others, and Green}{Ferguson et~al\mbox{.}}{2020}]%
        {ferguson_report_2020}
\bibfield{author}{\bibinfo{person}{Neil Ferguson}, \bibinfo{person}{Daniel
  Laydon}, \bibinfo{person}{Gemma Nedjati~Gilani}, \bibinfo{person}{Natsuko
  Imai}, \bibinfo{person}{Kylie Ainslie}, \bibinfo{person}{others}, {and}
  \bibinfo{person}{Green}.} \bibinfo{year}{2020}\natexlab{}.
\newblock \bibinfo{booktitle}{\emph{Report 9: {Impact} of non-pharmaceutical
  interventions ({NPIs}) to reduce {COVID19} mortality and healthcare demand}}.
\newblock \bibinfo{type}{{T}echnical {R}eport}.
\newblock


\bibitem[\protect\citeauthoryear{Flores}{Flores}{1989}]%
        {flores_utilization_1989}
\bibfield{author}{\bibinfo{person}{Benito~E. Flores}.}
  \bibinfo{year}{1989}\natexlab{}.
\newblock \showarticletitle{The utilization of the {Wilcoxon} test to compare
  forecasting methods: {A} note}.
\newblock \bibinfo{journal}{\emph{International Journal of Forecasting}}
  \bibinfo{volume}{5}, \bibinfo{number}{4} (\bibinfo{date}{Jan.}
  \bibinfo{year}{1989}), \bibinfo{pages}{529--535}.
\newblock


\bibitem[\protect\citeauthoryear{Grantz, Meredith, Cummings, Metcalf, Grenfell,
  et~al\mbox{.}}{Grantz et~al\mbox{.}}{2020}]%
        {grantz_use_2020}
\bibfield{author}{\bibinfo{person}{Kyra~H. Grantz}, \bibinfo{person}{Hannah~R.
  Meredith}, \bibinfo{person}{Derek A.~T. Cummings},
  \bibinfo{person}{C.~Jessica~E. Metcalf}, \bibinfo{person}{Bryan~T. Grenfell},
  {et~al\mbox{.}}} \bibinfo{year}{2020}\natexlab{}.
\newblock \showarticletitle{The use of mobile phone data to inform analysis of
  {COVID}-19 pandemic epidemiology}.
\newblock \bibinfo{journal}{\emph{Nature Communications}} \bibinfo{volume}{11},
  \bibinfo{number}{1} (\bibinfo{year}{2020}), \bibinfo{pages}{4961}.
\newblock


\bibitem[\protect\citeauthoryear{Hawkes and Oakes}{Hawkes and Oakes}{1974}]%
        {hawkes_cluster_1974}
\bibfield{author}{\bibinfo{person}{Alan~G. Hawkes} {and} \bibinfo{person}{David
  Oakes}.} \bibinfo{year}{1974}\natexlab{}.
\newblock \showarticletitle{A cluster process representation of a self-exciting
  process}.
\newblock \bibinfo{journal}{\emph{Journal of Applied Probability}}
  \bibinfo{volume}{11}, \bibinfo{number}{3} (\bibinfo{date}{Sept.}
  \bibinfo{year}{1974}), \bibinfo{pages}{493--503}.
\newblock


\bibitem[\protect\citeauthoryear{Kadar, Feuerriegel, Noulas, and Mascolo}{Kadar
  et~al\mbox{.}}{2020}]%
        {kadar_leveraging_2020}
\bibfield{author}{\bibinfo{person}{Cristina Kadar}, \bibinfo{person}{Stefan
  Feuerriegel}, \bibinfo{person}{Anastasios Noulas}, {and}
  \bibinfo{person}{Cecilia Mascolo}.} \bibinfo{year}{2020}\natexlab{}.
\newblock \showarticletitle{Leveraging {Mobility} {Flows} from {Location}
  {Technology} {Platforms} to {Test} {Crime} {Pattern} {Theory} in {Large}
  {Cities}}. In \bibinfo{booktitle}{\emph{{ICWSM}}}. \bibinfo{pages}{339--350}.
\newblock


\bibitem[\protect\citeauthoryear{Kafsi}{Kafsi}{2019}]%
        {kafsi_quantifying_2019}
\bibfield{author}{\bibinfo{person}{Mohamed Kafsi}.}
  \bibinfo{year}{2019}\natexlab{}.
\newblock \bibinfo{title}{Quantifying the {Accuracy} of {Mobility} {Insights}
  from {Cellular} {Network} {Data}}.
\newblock
\newblock
\urldef\tempurl%
\url{https://mkafsi.medium.com/quantifying-the-accuracy-of-mobility-insights-from-cellular-network-data-e5b83437a609}
\showURL{%
\tempurl}


\bibitem[\protect\citeauthoryear{Kapoor, Ben, Liu, Perozzi, Barnes,
  et~al\mbox{.}}{Kapoor et~al\mbox{.}}{2020}]%
        {kapoor_examining_2020}
\bibfield{author}{\bibinfo{person}{Amol Kapoor}, \bibinfo{person}{Xue Ben},
  \bibinfo{person}{Luyang Liu}, \bibinfo{person}{Bryan Perozzi},
  \bibinfo{person}{Matt Barnes}, {et~al\mbox{.}}}
  \bibinfo{year}{2020}\natexlab{}.
\newblock \showarticletitle{Examining {COVID}-19 {Forecasting} using
  {Spatio}-{Temporal} {Graph} {Neural} {Networks}}.
\newblock \bibinfo{journal}{\emph{arXiv preprint arXiv:2007.03113}}
  (\bibinfo{year}{2020}).
\newblock


\bibitem[\protect\citeauthoryear{Kelly, Park, Harrigan, Hoff, Lee, others, and
  Schoenberg}{Kelly et~al\mbox{.}}{2019}]%
        {kelly_real-time_2019}
\bibfield{author}{\bibinfo{person}{J.~Daniel Kelly}, \bibinfo{person}{Junhyung
  Park}, \bibinfo{person}{Ryan~J. Harrigan}, \bibinfo{person}{Nicole~A. Hoff},
  \bibinfo{person}{Sarita~D. Lee}, \bibinfo{person}{others}, {and}
  \bibinfo{person}{Schoenberg}.} \bibinfo{year}{2019}\natexlab{}.
\newblock \showarticletitle{Real-time predictions of the 2018–2019 {Ebola}
  virus disease outbreak in the {Democratic} {Republic} of the {Congo} using
  {Hawkes} point process models}.
\newblock \bibinfo{journal}{\emph{Epidemics}}  \bibinfo{volume}{28}
  (\bibinfo{year}{2019}), \bibinfo{pages}{100354}.
\newblock


\bibitem[\protect\citeauthoryear{Kermack and McKendrick}{Kermack and
  McKendrick}{1927}]%
        {kermack_contribution_1927}
\bibfield{author}{\bibinfo{person}{W.~O. Kermack} {and} \bibinfo{person}{A.~G.
  McKendrick}.} \bibinfo{year}{1927}\natexlab{}.
\newblock \showarticletitle{A contribution to the mathematical theory of
  epidemics}.
\newblock \bibinfo{journal}{\emph{Proceedings of the Royal Society of London.
  Series A, Containing Papers of a Mathematical and Physical Character}}
  \bibinfo{volume}{115}, \bibinfo{number}{772} (\bibinfo{date}{Aug.}
  \bibinfo{year}{1927}), \bibinfo{pages}{700--721}.
\newblock


\bibitem[\protect\citeauthoryear{Kim, Paini, and Jurdak}{Kim
  et~al\mbox{.}}{2019}]%
        {kim_modeling_2019}
\bibfield{author}{\bibinfo{person}{Minkyoung Kim}, \bibinfo{person}{Dean
  Paini}, {and} \bibinfo{person}{Raja Jurdak}.}
  \bibinfo{year}{2019}\natexlab{}.
\newblock \showarticletitle{Modeling stochastic processes in disease spread
  across a heterogeneous social system}.
\newblock \bibinfo{journal}{\emph{PNAS}} \bibinfo{volume}{116},
  \bibinfo{number}{2} (\bibinfo{year}{2019}), \bibinfo{pages}{401--406}.
\newblock


\bibitem[\protect\citeauthoryear{Lauer, Grantz, Bi, Jones, Zheng,
  et~al\mbox{.}}{Lauer et~al\mbox{.}}{2020}]%
        {lauer_incubation_2020}
\bibfield{author}{\bibinfo{person}{Stephen~A. Lauer}, \bibinfo{person}{Kyra~H.
  Grantz}, \bibinfo{person}{Qifang Bi}, \bibinfo{person}{Forrest~K. Jones},
  \bibinfo{person}{Qulu Zheng}, {et~al\mbox{.}}}
  \bibinfo{year}{2020}\natexlab{}.
\newblock \showarticletitle{The {Incubation} {Period} of {Coronavirus}
  {Disease} 2019 ({COVID}-19) {From} {Publicly} {Reported} {Confirmed} {Cases}:
  {Estimation} and {Application}}.
\newblock \bibinfo{journal}{\emph{Annals of Internal Medicine}}
  \bibinfo{volume}{172}, \bibinfo{number}{9} (\bibinfo{year}{2020}),
  \bibinfo{pages}{577--582}.
\newblock


\bibitem[\protect\citeauthoryear{Li, Pei, Chen, Song, Zhang, et~al\mbox{.}}{Li
  et~al\mbox{.}}{2020}]%
        {li_substantial_2020}
\bibfield{author}{\bibinfo{person}{Ruiyun Li}, \bibinfo{person}{Sen Pei},
  \bibinfo{person}{Bin Chen}, \bibinfo{person}{Yimeng Song},
  \bibinfo{person}{Tao Zhang}, {et~al\mbox{.}}}
  \bibinfo{year}{2020}\natexlab{}.
\newblock \showarticletitle{Substantial undocumented infection facilitates the
  rapid dissemination of novel coronavirus ({SARS}-{CoV}-2)}.
\newblock \bibinfo{journal}{\emph{Science}} \bibinfo{volume}{368},
  \bibinfo{number}{6490} (\bibinfo{year}{2020}), \bibinfo{pages}{489--493}.
\newblock


\bibitem[\protect\citeauthoryear{Liu, Gayle, Wilder-Smith, and Rocklöv}{Liu
  et~al\mbox{.}}{2020}]%
        {liu_reproductive_2020}
\bibfield{author}{\bibinfo{person}{Ying Liu}, \bibinfo{person}{Albert~A.
  Gayle}, \bibinfo{person}{Annelies Wilder-Smith}, {and}
  \bibinfo{person}{Joacim Rocklöv}.} \bibinfo{year}{2020}\natexlab{}.
\newblock \showarticletitle{The reproductive number of {COVID}-19 is higher
  compared to {SARS} coronavirus}.
\newblock \bibinfo{journal}{\emph{Journal of Travel Medicine}}
  \bibinfo{volume}{27}, \bibinfo{number}{2} (\bibinfo{year}{2020}).
\newblock


\bibitem[\protect\citeauthoryear{LLC}{LLC}{2020}]%
        {google_llc_google_2020}
\bibfield{author}{\bibinfo{person}{Google LLC}.}
  \bibinfo{year}{2020}\natexlab{}.
\newblock \bibinfo{title}{Google {COVID}-19 {Community} {Mobility} {Reports}.}
\newblock
\newblock
\urldef\tempurl%
\url{https://www.google.com/covid19/mobility/}
\showURL{%
\tempurl}


\bibitem[\protect\citeauthoryear{Miller, Foti, Lewnard, Jewell, Guestrin, and
  Fox}{Miller et~al\mbox{.}}{2020}]%
        {miller_mobility_2020}
\bibfield{author}{\bibinfo{person}{Andrew~C Miller},
  \bibinfo{person}{Nicholas~J Foti}, \bibinfo{person}{Joseph~A Lewnard},
  \bibinfo{person}{Nicholas~P Jewell}, \bibinfo{person}{Carlos Guestrin}, {and}
  \bibinfo{person}{Emily~B Fox}.} \bibinfo{year}{2020}\natexlab{}.
\newblock \showarticletitle{Mobility trends provide a leading indicator of
  changes in {SARS}-{CoV}-2 transmission}.
\newblock \bibinfo{journal}{\emph{medRxiv}} (\bibinfo{year}{2020}).
\newblock


\bibitem[\protect\citeauthoryear{Mohler, Schoenberg, Short, and Sledge}{Mohler
  et~al\mbox{.}}{2020}]%
        {mohler_analyzing_2020}
\bibfield{author}{\bibinfo{person}{George Mohler}, \bibinfo{person}{Frederic
  Schoenberg}, \bibinfo{person}{Martin~B Short}, {and} \bibinfo{person}{Daniel
  Sledge}.} \bibinfo{year}{2020}\natexlab{}.
\newblock \showarticletitle{Analyzing the {World}-{Wide} {Impact} of {Public}
  {Health} {Interventions} on the {Transmission} {Dynamics} of {COVID}-19}.
\newblock \bibinfo{journal}{\emph{arXiv preprint arXiv:2004.01714}}
  (\bibinfo{year}{2020}).
\newblock


\bibitem[\protect\citeauthoryear{of~Public~Health}{of~Public~Health}{2021}]%
        {federal_office_of_public_health_coronavirus_2021}
\bibfield{author}{\bibinfo{person}{Federal~Office of Public~Health}.}
  \bibinfo{year}{2021}\natexlab{}.
\newblock \bibinfo{title}{Coronavirus: {Situation} in {Switzerland}}.
\newblock
\newblock
\urldef\tempurl%
\url{https://www.bag.admin.ch/bag/en/home/krankheiten/ausbrueche-epidemien-pandemien/aktuelle-ausbrueche-epidemien/novel-cov/situation-schweiz-und-international.html}
\showURL{%
\tempurl}


\bibitem[\protect\citeauthoryear{Office}{Office}{2021}]%
        {office_population_2021}
\bibfield{author}{\bibinfo{person}{Federal~Statistical Office}.}
  \bibinfo{year}{2021}\natexlab{}.
\newblock \bibinfo{title}{Population}.
\newblock
\newblock
\urldef\tempurl%
\url{https://www.bfs.admin.ch/bfs/en/home/statistiken/bevoelkerung.html}
\showURL{%
\tempurl}


\bibitem[\protect\citeauthoryear{Ogata}{Ogata}{1988}]%
        {ogata_statistical_1988}
\bibfield{author}{\bibinfo{person}{Yosihiko Ogata}.}
  \bibinfo{year}{1988}\natexlab{}.
\newblock \showarticletitle{Statistical {Models} for {Earthquake} {Occurrences}
  and {Residual} {Analysis} for {Point} {Processes}}.
\newblock \bibinfo{journal}{\emph{J. Amer. Statist. Assoc.}}
  \bibinfo{volume}{83}, \bibinfo{number}{401} (\bibinfo{date}{March}
  \bibinfo{year}{1988}), \bibinfo{pages}{9--27}.
\newblock


\bibitem[\protect\citeauthoryear{Oliver, Lepri, Sterly, Lambiotte, Deletaille,
  et~al\mbox{.}}{Oliver et~al\mbox{.}}{2020}]%
        {oliver_mobile_2020}
\bibfield{author}{\bibinfo{person}{Nuria Oliver}, \bibinfo{person}{Bruno
  Lepri}, \bibinfo{person}{Harald Sterly}, \bibinfo{person}{Renaud Lambiotte},
  \bibinfo{person}{Sébastien Deletaille}, {et~al\mbox{.}}}
  \bibinfo{year}{2020}\natexlab{}.
\newblock \showarticletitle{Mobile phone data for informing public health
  actions across the {COVID}-19 pandemic life cycle}.
\newblock \bibinfo{journal}{\emph{Science advances}} \bibinfo{volume}{6},
  \bibinfo{number}{23} (\bibinfo{year}{2020}).
\newblock


\bibitem[\protect\citeauthoryear{Pei, Kandula, and Shaman}{Pei
  et~al\mbox{.}}{2020}]%
        {pei_differential_2020}
\bibfield{author}{\bibinfo{person}{Sen Pei}, \bibinfo{person}{Sasikiran
  Kandula}, {and} \bibinfo{person}{Jeffrey Shaman}.}
  \bibinfo{year}{2020}\natexlab{}.
\newblock \showarticletitle{Differential {Effects} of {Intervention} {Timing}
  on {COVID}-19 {Spread} in the {United} {States}}.
\newblock \bibinfo{journal}{\emph{Science advances}} \bibinfo{volume}{6},
  \bibinfo{number}{49} (\bibinfo{year}{2020}).
\newblock


\bibitem[\protect\citeauthoryear{Persson, Parie, and Feuerriegel}{Persson
  et~al\mbox{.}}{2021}]%
        {persson_monitoring_2021}
\bibfield{author}{\bibinfo{person}{Joel Persson}, \bibinfo{person}{Jurriaan~F.
  Parie}, {and} \bibinfo{person}{Stefan Feuerriegel}.}
  \bibinfo{year}{2021}\natexlab{}.
\newblock \showarticletitle{Monitoring the {COVID}-19 epidemic with nationwide
  telecommunication data}.
\newblock \bibinfo{journal}{\emph{arXiv preprint arXiv:2101.02521}}
  (\bibinfo{year}{2021}).
\newblock


\bibitem[\protect\citeauthoryear{Qian and Alaa}{Qian and Alaa}{2020}]%
        {qian_when_2020}
\bibfield{author}{\bibinfo{person}{Zhaozhi Qian} {and} \bibinfo{person}{Ahmed~M
  Alaa}.} \bibinfo{year}{2020}\natexlab{}.
\newblock \showarticletitle{When and {How} to {Lift} the {Lockdown}? {Global}
  {COVID}-19 {Scenario} {Analysis} and {Policy} {Assessment} using
  {Compartmental} {Gaussian} {Processes}}.
\newblock \bibinfo{journal}{\emph{NeurIPS}}  \bibinfo{volume}{33}
  (\bibinfo{year}{2020}).
\newblock


\bibitem[\protect\citeauthoryear{Rizoiu, Mishra, Kong, Carman, and Xie}{Rizoiu
  et~al\mbox{.}}{2018}]%
        {rizoiu_sir-hawkes_2018}
\bibfield{author}{\bibinfo{person}{Marian-Andrei Rizoiu},
  \bibinfo{person}{Swapnil Mishra}, \bibinfo{person}{Quyu Kong},
  \bibinfo{person}{Mark Carman}, {and} \bibinfo{person}{Lexing Xie}.}
  \bibinfo{year}{2018}\natexlab{}.
\newblock \showarticletitle{{SIR}-{Hawkes}: {Linking} {Epidemic} {Models} and
  {Hawkes} {Processes} to {Model} {Diffusions} in {Finite} {Populations}}.
\newblock \bibinfo{journal}{\emph{WWW}} (\bibinfo{year}{2018}),
  \bibinfo{pages}{419--428}.
\newblock


\bibitem[\protect\citeauthoryear{Rostami-Tabar and
  Rendon-Sanchez}{Rostami-Tabar and Rendon-Sanchez}{2021}]%
        {rostami-tabar_forecasting_2021}
\bibfield{author}{\bibinfo{person}{Bahman Rostami-Tabar} {and}
  \bibinfo{person}{Juan~F. Rendon-Sanchez}.} \bibinfo{year}{2021}\natexlab{}.
\newblock \showarticletitle{Forecasting {COVID}-19 daily cases using phone call
  data}.
\newblock \bibinfo{journal}{\emph{Applied soft computing}}
  \bibinfo{volume}{100} (\bibinfo{year}{2021}), \bibinfo{pages}{106932}.
\newblock


\bibitem[\protect\citeauthoryear{Schlickeiser and Schlickeiser}{Schlickeiser
  and Schlickeiser}{2020}]%
        {schlickeiser_gaussian_2020}
\bibfield{author}{\bibinfo{person}{Reinhard Schlickeiser} {and}
  \bibinfo{person}{Frank Schlickeiser}.} \bibinfo{year}{2020}\natexlab{}.
\newblock \showarticletitle{A {Gaussian} {Model} for the {Time} {Development}
  of the {Sars}-{Cov}-2 {Corona} {Pandemic} {Disease}. {Predictions} for
  {Germany} {Made} on 30 {March} 2020}.
\newblock \bibinfo{journal}{\emph{Physics}} \bibinfo{volume}{2},
  \bibinfo{number}{2} (\bibinfo{year}{2020}), \bibinfo{pages}{164--170}.
\newblock


\bibitem[\protect\citeauthoryear{Schoenberg, Hoffmann, and Harrigan}{Schoenberg
  et~al\mbox{.}}{2019}]%
        {schoenberg_recursive_2019}
\bibfield{author}{\bibinfo{person}{Frederic Schoenberg}, \bibinfo{person}{Marc
  Hoffmann}, {and} \bibinfo{person}{Ryan Harrigan}.}
  \bibinfo{year}{2019}\natexlab{}.
\newblock \showarticletitle{A recursive point process model for infectious
  diseases}.
\newblock \bibinfo{journal}{\emph{Ann. Inst. Stat.}} \bibinfo{volume}{71},
  \bibinfo{number}{5} (\bibinfo{year}{2019}), \bibinfo{pages}{1271--1287}.
\newblock


\bibitem[\protect\citeauthoryear{Swisscom}{Swisscom}{2021}]%
        {swisscom_swisscom_2021}
\bibfield{author}{\bibinfo{person}{Swisscom}.} \bibinfo{year}{2021}\natexlab{}.
\newblock \bibinfo{title}{Swisscom {Mobility} {Insights}}.
\newblock
\newblock
\urldef\tempurl%
\url{https://www.swisscom.ch/en/business/enterprise/offer/enterprise-mobile/mobility-insights.html}
\showURL{%
\tempurl}


\bibitem[\protect\citeauthoryear{Tibshirani}{Tibshirani}{1996}]%
        {tibshirani_regression_1996}
\bibfield{author}{\bibinfo{person}{Robert Tibshirani}.}
  \bibinfo{year}{1996}\natexlab{}.
\newblock \showarticletitle{Regression {Shrinkage} and {Selection} {Via} the
  {Lasso}}.
\newblock \bibinfo{journal}{\emph{J R Stat Soc Series B Stat Methodol}}
  \bibinfo{volume}{58}, \bibinfo{number}{1} (\bibinfo{year}{1996}),
  \bibinfo{pages}{267--288}.
\newblock


\bibitem[\protect\citeauthoryear{Wang, Zhou, Mascolo, Noulas, Xie,
  et~al\mbox{.}}{Wang et~al\mbox{.}}{2018}]%
        {wang_predicting_2018}
\bibfield{author}{\bibinfo{person}{Yingzi Wang}, \bibinfo{person}{Xiao Zhou},
  \bibinfo{person}{Cecilia Mascolo}, \bibinfo{person}{Anastasios Noulas},
  \bibinfo{person}{Xing Xie}, {et~al\mbox{.}}} \bibinfo{year}{2018}\natexlab{}.
\newblock \showarticletitle{Predicting the {Spatio}-{Temporal} {Evolution} of
  {Chronic} {Diseases} in {Population} with {Human} {Mobility} {Data}.}. In
  \bibinfo{booktitle}{\emph{{IJCAI}}}.
\newblock


\bibitem[\protect\citeauthoryear{WHO}{WHO}{2021}]%
        {who_coronavirus_2021}
\bibfield{author}{\bibinfo{person}{WHO}.} \bibinfo{year}{2021}\natexlab{}.
\newblock \bibinfo{title}{Coronavirus {Disease} ({COVID}-19) {Situation}
  {Reports}}.
\newblock
\newblock
\urldef\tempurl%
\url{https://www.who.int/emergencies/diseases/novel-coronavirus-2019/situation-reports}
\showURL{%
\tempurl}


\bibitem[\protect\citeauthoryear{Wieczorek, Siłka, and Woźniak}{Wieczorek
  et~al\mbox{.}}{2020}]%
        {wieczorek_neural_2020}
\bibfield{author}{\bibinfo{person}{Michał Wieczorek}, \bibinfo{person}{Jakub
  Siłka}, {and} \bibinfo{person}{Marcin Woźniak}.}
  \bibinfo{year}{2020}\natexlab{}.
\newblock \showarticletitle{Neural network powered {COVID}-19 spread
  forecasting model}.
\newblock \bibinfo{journal}{\emph{Chaos, Solitons \& Fractals}}
  \bibinfo{volume}{140} (\bibinfo{year}{2020}), \bibinfo{pages}{110203}.
\newblock


\bibitem[\protect\citeauthoryear{Winkelmann}{Winkelmann}{2008}]%
        {winkelmann_econometric_2008}
\bibfield{author}{\bibinfo{person}{Rainer Winkelmann}.}
  \bibinfo{year}{2008}\natexlab{}.
\newblock \bibinfo{booktitle}{\emph{Econometric {Analysis} of {Count} {Data}}}.
\newblock \bibinfo{publisher}{Springer Science \& Business Media}.
\newblock


\bibitem[\protect\citeauthoryear{Xiong, Hu, Yang, Luo, and Zhang}{Xiong
  et~al\mbox{.}}{2020}]%
        {xiong_mobile_2020}
\bibfield{author}{\bibinfo{person}{Chenfeng Xiong}, \bibinfo{person}{Songhua
  Hu}, \bibinfo{person}{Mofeng Yang}, \bibinfo{person}{Weiyu Luo}, {and}
  \bibinfo{person}{Lei Zhang}.} \bibinfo{year}{2020}\natexlab{}.
\newblock \showarticletitle{Mobile device data reveal the dynamics in a
  positive relationship between human mobility and {COVID}-19 infections}.
\newblock \bibinfo{journal}{\emph{PNAS}} \bibinfo{volume}{117},
  \bibinfo{number}{44} (\bibinfo{year}{2020}), \bibinfo{pages}{27087--27089}.
\newblock


\bibitem[\protect\citeauthoryear{You, Deng, Hu, Sun, Lin, et~al\mbox{.}}{You
  et~al\mbox{.}}{2020}]%
        {you_estimation_2020}
\bibfield{author}{\bibinfo{person}{Chong You}, \bibinfo{person}{Yuhao Deng},
  \bibinfo{person}{Wenjie Hu}, \bibinfo{person}{Jiarui Sun},
  \bibinfo{person}{Qiushi Lin}, {et~al\mbox{.}}}
  \bibinfo{year}{2020}\natexlab{}.
\newblock \showarticletitle{Estimation of the time-varying reproduction number
  of {COVID}-19 outbreak in {China}}.
\newblock \bibinfo{journal}{\emph{International Journal of Hygiene and
  Environmental Health}}  \bibinfo{volume}{228} (\bibinfo{year}{2020}),
  \bibinfo{pages}{113555}.
\newblock


\bibitem[\protect\citeauthoryear{Zhao, Li, Liu, Zhu, Ma, et~al\mbox{.}}{Zhao
  et~al\mbox{.}}{2020}]%
        {zhao_prediction_2020}
\bibfield{author}{\bibinfo{person}{Zebin Zhao}, \bibinfo{person}{Xin Li},
  \bibinfo{person}{Feng Liu}, \bibinfo{person}{Gaofeng Zhu},
  \bibinfo{person}{Chunfeng Ma}, {et~al\mbox{.}}}
  \bibinfo{year}{2020}\natexlab{}.
\newblock \showarticletitle{Prediction of the {COVID}-19 spread in {African}
  countries and implications for prevention and control: {A} case study in
  {South} {Africa}, {Egypt}, {Algeria}, {Nigeria}, {Senegal} and {Kenya}}.
\newblock \bibinfo{journal}{\emph{Science of The Total Environment}}
  \bibinfo{volume}{729} (\bibinfo{year}{2020}), \bibinfo{pages}{138959}.
\newblock


\bibitem[\protect\citeauthoryear{Zhou, Hristova, Noulas, Mascolo, and
  Sklar}{Zhou et~al\mbox{.}}{2017}]%
        {zhou_cultural_2017}
\bibfield{author}{\bibinfo{person}{Xiao Zhou}, \bibinfo{person}{Desislava
  Hristova}, \bibinfo{person}{Anastasios Noulas}, \bibinfo{person}{Cecilia
  Mascolo}, {and} \bibinfo{person}{Max Sklar}.}
  \bibinfo{year}{2017}\natexlab{}.
\newblock \showarticletitle{Cultural investment and urban socio-economic
  development: a geosocial network approach}.
\newblock \bibinfo{journal}{\emph{Royal Society Open Science}}
  \bibinfo{volume}{4}, \bibinfo{number}{9} (\bibinfo{year}{2017}),
  \bibinfo{pages}{170413}.
\newblock


\end{thebibliography}

\end{document}